\documentclass[apj]{emulateapj}
 
\renewcommand\email\texttt

\newcommand\msun{{{\rm M}_{\odot}}}
\newcommand\Msun{{{\rm M}_{\odot}}}
\newcommand\kms{{\rm kms}^{-1}}
  
\def\spose#1{\hbox to 0pt{#1\hss}}
\def\lta{\mathrel{\spose{\lower 3pt\hbox{$\sim$}}
    \raise 2.0pt\hbox{$<$}}}
\def\gta{\mathrel{\spose{\lower 3pt\hbox{$\sim$}}
    \raise 2.0pt\hbox{$>$}}}
\shorttitle{The Halo Shape}
\shortauthors{Fellhauer et al.}

\begin{document}

\title{The Origin of the Bifurcation in the Sagittarius Stream} 
\author{
M. Fellhauer\altaffilmark{1},
V. Belokurov\altaffilmark{1},
N. W. Evans\altaffilmark{1},
M. I. Wilkinson\altaffilmark{1},
D. B. Zucker\altaffilmark{1},
G. Gilmore\altaffilmark{1},
M. J. Irwin\altaffilmark{1},
D. M. Bramich\altaffilmark{1},
S. Vidrih\altaffilmark{1},
R. F. G. Wyse\altaffilmark{2},
T. C. Beers\altaffilmark{3},
J. Brinkmann\altaffilmark{4}.
}
\altaffiltext{1}{Institute of Astronomy, University of Cambridge,
Madingley Road, Cambridge CB3 0HA, UK;\email{madf,vasily,nwe@ast.cam.ac.uk}}
\altaffiltext{2}{The Johns Hopkins University, 3701 San Martin Drive,
Baltimore, MD 21218}
\altaffiltext{3}{Department of Physics and Astronomy, CSCE: Center for
the Study of Cosmic Evolution, and JINA: Joint Institute for Nuclear
Astrophysics, Michigan State University, East Lansing, MI 48824}
\altaffiltext{4}{Apache Point Observatory, P.O. Box 59, Sunspot, NM
  88349}

\begin{abstract}
The latest Sloan Digital Sky Survey data reveal a prominent
bifurcation in the distribution of debris of the Sagittarius dwarf
spheroidal (Sgr) beginning at a right ascension of $\alpha \approx
190^\circ$. Two branches of the stream (A and B) persist at roughly
the same heliocentric distance over at least $50^\circ$ of arc. There
is also evidence for a more distant structure (C) well behind the A
branch.  This paper provides the first explanation for the
bifurcation. It is caused by the projection of the young leading (A)
and old trailing (B) tidal arms of the Sgr, whilst the old leading arm
(C) lies well behind A. This explanation is only possible if the halo
is close to spherical, as the angular difference between the branches
is a measure of the precession of the orbital plane.
\end{abstract}

\maketitle

\section{Introduction}
\label{sec:intro}

The disrupting Sagittarius dwarf spheroidal galaxy (Sgr) was
discovered by Ibata, Gilmore \& Irwin in 1994. It was soon realized
that the Sgr provided a powerful tool for the study of the
Galaxy~\citep{Ib97}.  The nucleus of the Sgr has survived for many
orbits around the Galaxy, whilst its tidal tails have now been
detected over a full $360^\circ$ on the Sky \citep[see
e.g.][]{To98,Ma03,Be06}. The disrupted fragments of the Sgr diffuse in
the Galactic potential. As the pericenter of the Sgr's orbit is $\sim
16$ kpc, whilst its apocenter is $\sim 60$ kpc, the debris provides a
strong constraint on the Galaxy's halo.

\begin{figure*}
\begin{center}
\includegraphics[height=10cm]{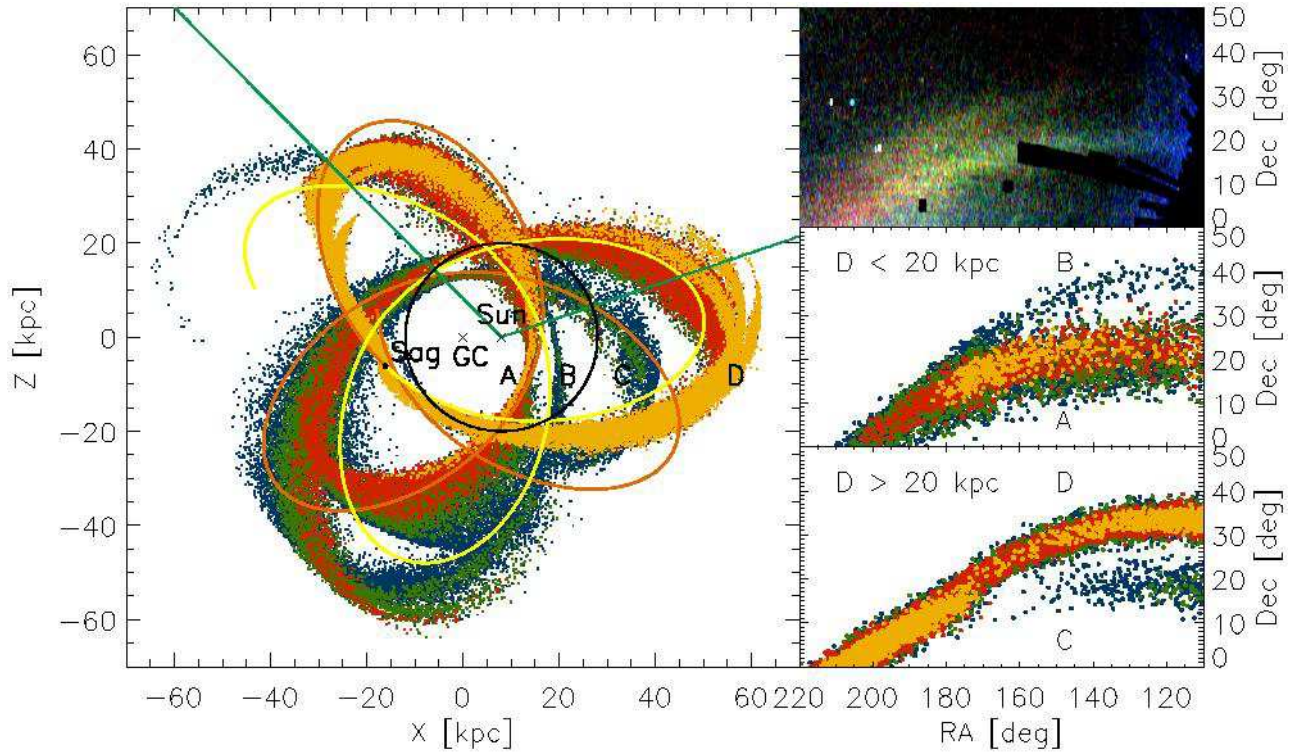} 
\caption{Left: Simulation showing the tails of the Sgr
dSph. Particles are color-coded according to when they were lost
(gold: $< 4$~Gyr ago, red: between $4$ and $5.7$~Gyr, green: between
$5.7$ and $7.4$~Gyr and blue $>7.4$~Gyr ago). The yellow (orange)
curves show the past (future) behavior of the Sgr's orbit over 2 Gyr.
The orbital period is 0.7 Gyr. The positions of the Galactic centre
(GC), the Sun, and Sagittarius (Sag) are marked.  The green lines show
the right ascension range $110^\circ < \alpha < 220^\circ$, which
corresponds to the SDSS data analyzed by Belokurov et al. (2006). The
4 streams are marked A (young leading arm), B (old trailing arm), C
(old leading arm) and D (young trailing arm).  The circle gives the
distance cut-off at 20 kpc.  Upper Right: The SDSS data from Belokurov
et al. (2006), with stars color-coded according to magnitude.  Middle
and Lower Right: Scatter plots in right ascension and declination of
the tidal debris.  Only particles within (beyond) a heliocentric
distance of $20$ kpc are plotted in the middle (lower) panels. The
black squares show the field locations of Belokurov et al. (2006).  In
the middle panel, streams A and B are clearly visible. The upper arm
is the old trailing material, while the lower arm is the young leading
material.  In the lower panel, the old leading material is in the
lower and the young trailing material in the upper branch. [The
simulation uses a Miyamoto-Nagai disk and logarithmic halo with
$q_\phi = 1.05$, together with the set d of proper motions. The mass
of Sgr is $10^8 M_\odot$]}
\label{fig:xz}
\end{center}
\end{figure*}

The morphology of the Sgr stream is known in detail in the Galactic
southern hemisphere thanks to 2MASS \citep{Ma03,Ma04,Sk06}.
\citet{Ne02,Ne03} made early detections of Sgr tidal debris in data
from the Sloan Digital Sky Survey \citep[SDSS, see][]{Ho01,St02,Sm02,
Pi03,Iv04,Gu06}. Then, \citet{Be06} used a color cut to pick out the
upper main sequence and turn-off stars in SDSS Data Release 5
\citep[DR5,][]{Am06} belonging to the stream. In the Galactic northern
hemisphere, they found a prominent bifurcation or branching in the
stream, beginning at a right ascension $\alpha \approx 190^{\circ}$
(see the upper right panel of Figure~\ref{fig:xz}). The lower and
upper declination branches of the stream, labelled A and B, can be
traced until right ascensions of at least $\alpha \approx
140^{\circ}$. Using the location of the subgiant branch as an
estimator, the A and B branches are reckoned to be at similar
distances.  There is also evidence in the data of a fainter, still
more distant stream (C) directly behind the A branch.

Belokurov et al.'s (2006) dataset is important for two reasons. First,
it traces the Sgr stream around the North Galactic Cap, the very spot
at which oblate and prolate dark halos give different
predictions~\citep[e.g.,][]{He04a}.  Second, the debris in 2MASS is
dynamically younger than that found in SDSS, so the SDSS data should
give stronger constraints, as the stars have had longer to move in the
Galactic potential.

Early explorations of the evolution of the Sgr suggested that the
Galactic halo may be close to spherical~\citep{Ib01b,Ma03}. For
example, \citet{Jo05} showed that the precession apparent in Sgr
debris in the 2MASS dataset strongly favored mildly oblate halos.
However, \citet{He04a} pointed out that many of the earlier datasets
are restricted to stars that have only recently been torn off the Sgr
and so have not diffused in the Galactic potential. In fact,
\citet{He04b} argued that the velocity measurements of 2MASS M giants
in the leading arm favor strongly prolate halos and this was
subsequently confirmed by \citet{La05}. Hence, present studies of the
disruption of the Sgr have reached an impasse, with different datasets
pointing to dramatically different flattenings.

Here, our aim is to show how the newly discovered bifurcation
arises. Using numerical simulations, we argue that the complex
morphology of the Sgr stream uncovered by Belokurov et al. (2006) can
only be reproduced if the Galaxy's halo is close to spherical. Although
our models do not resolve the contradiction between the precession
rate and the velocities in the leading arm, they do provide a new and
powerful argument in favor of an almost spherical Galaxy halo.

\begin{table*}
\begin{center}
\caption{\label{tab:pmdata} Sets of proper motions used in the simulations.}
\begin{tabular}{cccccc}
\hline
Label & $\mu_{\alpha}\cos\delta$ & $\mu_{\delta}$ & $v_{\rm rad}$ &
Remarks \\  
\null & (in mas\,yr$^{-1}$) & (in mas\,yr$^{-1}$) & (in km\,s$^{-1}$) 
& \null\\ 
\hline
a & -2.65 & -0.88 & 137 & HST measurement (Ibata et al 2001b) \\
b & -2.8 & -1.4 & 137   & Schmidt plates measurement (Irwin et
al. 1996) \\
c & -2.9  & -1.5  & 137 & $< 1 \sigma$ variation of Schmidt plates \\ 
d & -3.02 & -1.49 & 137 & Simulation fit from Law et al. (2005) \\
e & -3.05 & 1.28 & 137  & $5\sigma$ variation of HST values \\\hline
\end{tabular}
\end{center}
\end{table*}
\begin{figure*}
\begin{center}
\includegraphics[height=10cm]{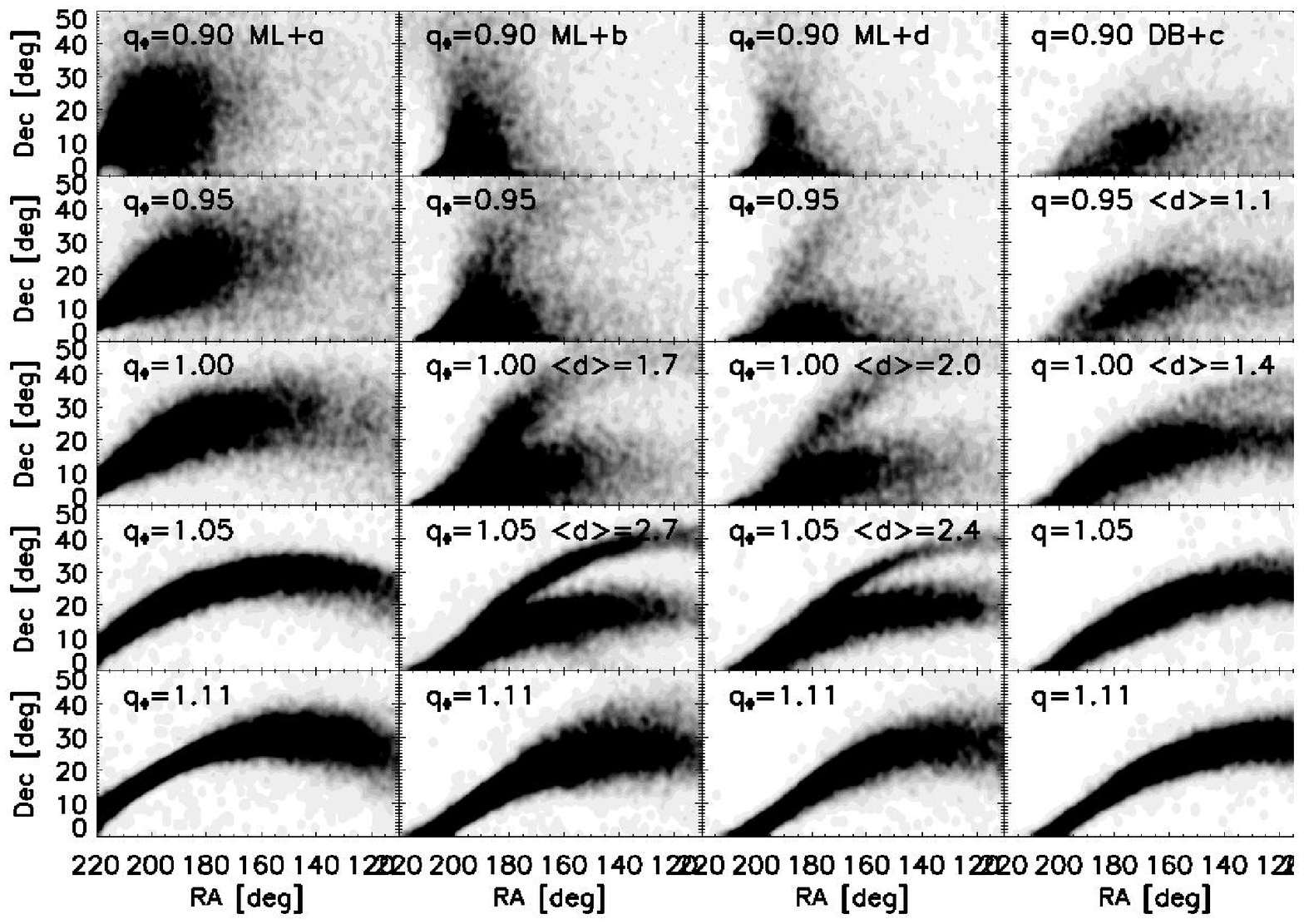}
\caption{\label{fig:sims} The projected density of the young leading
and old trailing tidal debris of the Sgr in a sequence of simulations
with different halo flattenings, proper motions and Galaxy models. The
key at the top of each column has the following significance: ML means
Miyamoto-Nagai disk and logarithmic halo models, DB means the Dehnen
\& Binney models, whilst a, b, c, d or e refers to the choice of
proper motions. The panels in a given column differ only in the
flattening of the halo, with $q_\phi$ for the ML models and $q$ for
the DB models recorded in the upper left corner. The mass of the Sgr
is $10^8 M_\odot$. Note that the particles in the simulation represent
both stars and dark matter.}
\end{center}
\end{figure*}

\section{The Significance of the Bifurcation}

The results of a typical simulation of the tidal disruption of the Sgr
in a nearly spherical potential are shown in Fig.~\ref{fig:xz}. We
will give the details of the simulation set-up shortly, but at the
moment our aim is to gain a qualitative understanding of why the
bifurcation occurs. The particles in Fig.~\ref{fig:xz} are color-coded
according to when they were torn off the Sgr. In the direction of the
SDSS DR5 data -- namely, the opening angle defined by the green lines
-- there are 4 distinct streams of material. They are the young
leading arm (labelled A), the old trailing arm (B), the old leading
arm (C) and the young trailing arm (D). Here, old and young indicate
when the stars were torn off. 3 out of the 4 streams are identified in
the SDSS DR5 dataset. Stars belonging to the D stream are more
difficult to detect as they occur in DR5 primarily in the range
$180^\circ \lta \alpha \lta 220^\circ$, where they are not easy to
untangle from the other streams. The simulation data are separated
according to heliocentric distance and then projected onto the sky as
viewed from the Sun, as shown in the lower right panels of
Fig.~\ref{fig:xz}. The young leading arm provides branch A and the old
trailing arm branch B of the bifurcated stream of Belokurov et
al. (2006). These two narrow branches are at similar heliocentric
distances, as required to match the data.  The material in these
branches is about two revolutions apart in orbital phase.  For the
material beyond 20 kpc, the older leading material is in the lower
declination branch, while the younger trailing material is in the
upper. The old leading arm (C) provides the more distant and fainter
stream detected by Belokurov et al. (2006) behind the A branch.  The
young trailing arm (D) lies behind the B branch and so is not the
structure seen by Belokurov et al. (2006).

The Sun lies roughly in the orbital plane of the Sgr. If the potential
were exactly spherical, the debris of the Sgr would lie in a single
plane and no bifurcation would exist. Any asphericity (whether
intrinsic to the halo or produced by the bulge and disk) causes the
orbital plane to precess and therefore the planes of the 4 arms to be
slightly different. The positional difference between branches A and B
is a direct measure of the precession over two orbital revolutions and
hence the asphericity of the potential. The facts that (i) branches A
and B are so close in projection and (ii) branch C lies behind branch
A suggest that the precession is small, and that the potential is
close to spherical. If the halo is too oblate or prolate, then debris
is scattered over a wide range of locations and does not lie in thin,
almost overlapping streams on the sky.  To back up this qualitative
argument, let us now describe a suite of simulations developed to
measure the properties of the bifurcation as a function of halo
flattening, Sgr mass and proper motion.

\begin{table}
\begin{center}
\caption{\label{tab:simres} The Strength and Range of the Bifurcation
  in Nearly Spherical Haloes}
\begin{tabular}{|c|c|c|c|c|c|c|}
\multicolumn{7}{c}{Logarithmic Halo\tablenotemark{a}}\\
\hline
 $Q_\phi$ & $q_\phi$  & a & b & c & d & e \\\hline
    0.92     &1.0  & -- &1.7 & 1.9 &2.0 & 1.5 \\
             &  & -- &$170^\circ$ & $180^\circ$ &
$190^\circ$ & $160^\circ$ \\ \hline
     0.95  & 1.05 & -- &2.7 &2.3 & 2.4 & 1.7 \\
       &  & -- &$190^\circ$ & $190^\circ$ & $170^\circ$ &
     $150^\circ$ \\ \hline
\end{tabular}
\tablenotetext{a}{All Galactic models with $q_\phi$ = 0.8, 0.9, 0.95,
  1.0, 1.05, 1.11, 1.25 and 1.5 were investigated. If there is no
  bifurcation, or if the lower branch of the bifurcation bends back to
  negative declinations, the model is discarded.  For models in which
  there is a bifurcation, the strength $\langle d \rangle$, and the onset
  $\alpha_o$  are given for each set of proper motions.}

\begin{tabular}{|c|c|c|c|c|c|c|}
\multicolumn{7}{c}{Dehnen \& Binney Models\tablenotemark{b}}\\
\hline
   $Q_\phi$ & $q$ & a & b & c & d & e \\ \hline
    0.95 & 0.95 & -- & 1.1 & 1.1 & -- & -- \\
         &     & --    & $160^\circ$ & $150^\circ$ & -- & -- \\\hline
    0.97  & 1.0  & -- &1.4 &1.4 & -- & -- \\     
          &   & -- & $130^\circ$& $150^\circ$& -- & \\ \hline
\end{tabular}
\tablenotetext{b}{All Galactic models with $q$ = 0.8, 0.9, 0.95, 1.0, 1.05,
  1.11, 1.25 and 1.5 were investigated but only those with
  bifurcations are reported.}
\end{center}
\end{table}
\begin{figure}
\begin{center}
\includegraphics[height=10cm]{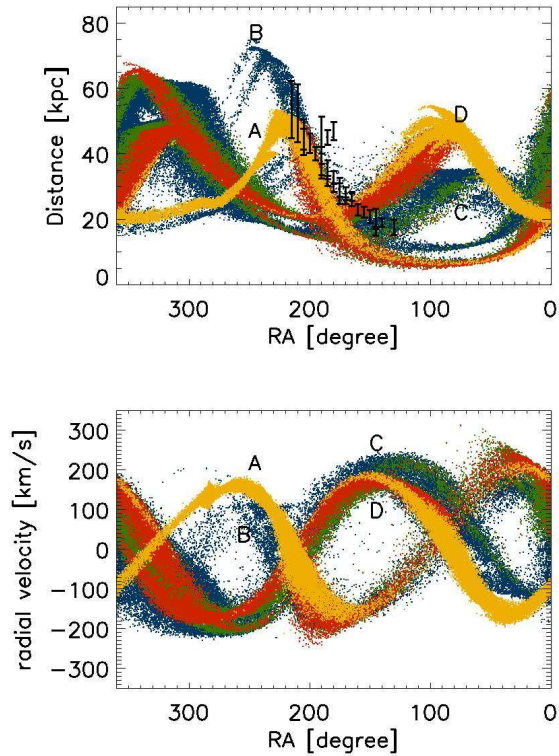}
\caption{\label{fig:error} Plots of the heliocentric distance and
velocity versus right ascension for the same model as in
Figure~\ref{fig:xz}. Again, particles are color-coded according to
when they were lost (gold: $< 4$~Gyr ago, red: between $4$ and
$5.7$~Gyr, green: between $5.7$ and $7.4$~Gyr and blue $>7.4$~Gyr
ago). The datapoints give the heliocentric distances to the streams,
as derived from fitting the subgiant branch described in Belokurov et
al. (2006). Note that the distances to the A and B streams are too
small for right ascensions $\alpha \lta 190^\circ$.}
\end{center}
\end{figure}
\begin{figure*}
\begin{center}
\includegraphics[height=10cm]{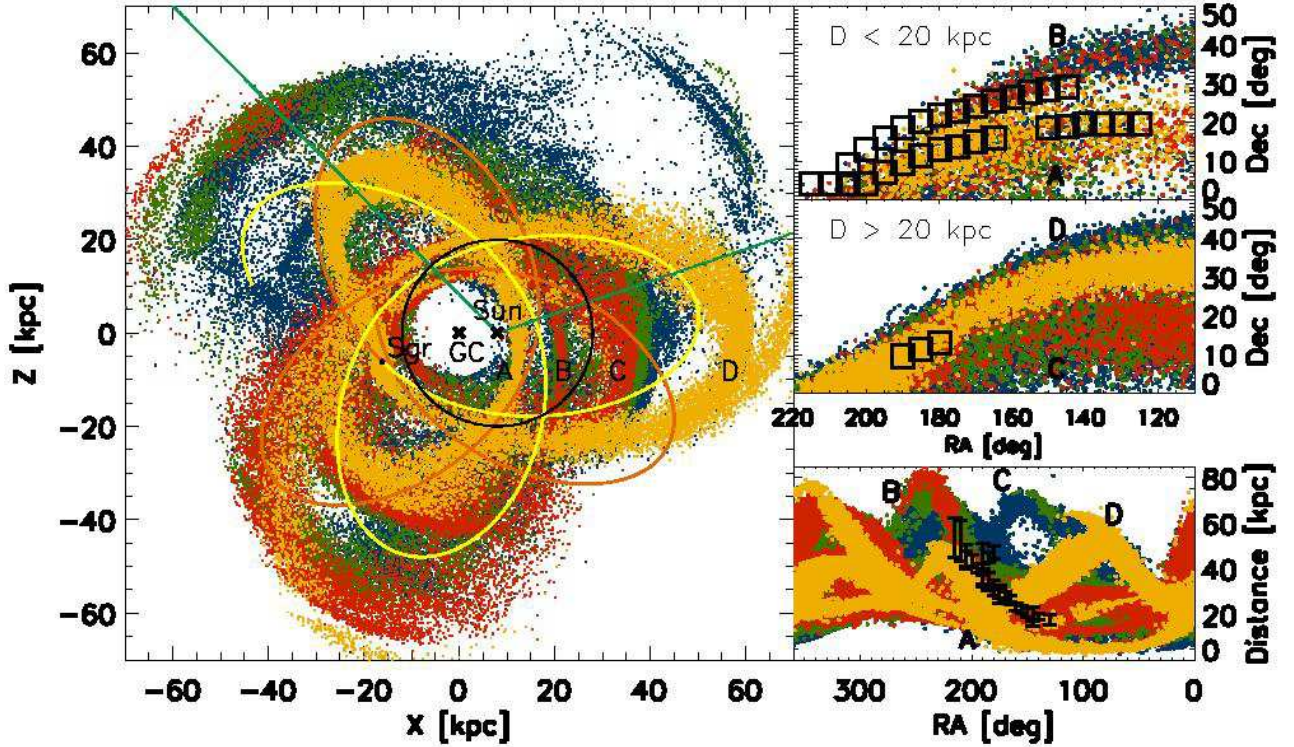}
\caption{\label{fig:moremass} As Figure 1, but now the mass of the Sgr
has been increased to $5 \times 10^8 M_\odot$. A bifurcation is
visible in the right-hand panels, but it is less dramatic than in the
data.}
\end{center}
\end{figure*}

\section{Simulations}

\subsection{Set-up}

The present position of the Sgr dSph is ($\alpha, \delta) =
(283.7^\circ, -30.5^\circ$), while its heliocentric distance is
$25\pm2$ kpc and radial velocity is $137\ \kms$ ~\citep{Ib97}.  Listed
in Table~\ref{tab:pmdata} are two measurements of the proper motion of
the Sgr, the first from~\citet{Ir96} using Schmidt plates and the
second from~\citet{Ib01b} using HST data.  Dinescu et al.'s (2005)
recent measurement agrees with that of \citet{Ir96} within the errors.
Given a choice of proper motions, we integrate back in time for $10$
Gyr adopting a potential for the Galaxy. At the final position, we
insert a Plummer sphere containing $10^6$ particles with a scalelength
of $350$~pc. We investigate models with a total mass of between $10^8$
and $10^9$ M$_{\odot}$.  The particles are integrated forward using
the particle-mesh N-Body code {\sc Superbox} \citep{Fe00} until the
position today is reached again.

While the present position of the Sgr is unchanged in all our
simulations, the range of proper motions recorded in
Table~\ref{tab:pmdata} is investigated. For the Galactic potential, we
use one of two possibilities.  In the first (denoted by ML), the halo
is represented by a logarithmic potential of the form
\begin{eqnarray}
  \label{eq:halopot}
  \Phi_{\rm halo}(r) & = & \frac{v_{0}^{2}}{2} \ln \left( R^{2} +
    z^2 q_\Phi^{-2} +  d^{2} \right),
\end{eqnarray}
with $v_{0}=186$~km\,s$^{-1}$ and $d=12$~kpc (where $R$ and $z$ are
cylindrical coordinates). The parameter $q_{\Phi}$ is the axis ratio
of the equipotentials. It controls whether the halo is spherical
($q_\Phi=1$), oblate ($q_\Phi<1$) or prolate ($q_\Phi>1$).  In
general, $q_\Phi$ is of course not the same as the axis ratio in the
density $q$, which varies with radius for the logarithmic potential
(see e.g., Evans 1993).  The disc is represented by a Miyamoto-Nagai
potential:
\begin{eqnarray}  \label{eq:discpot}
  \Phi_{\rm disc}(R,z) & = & \frac{G M_{\rm d}} { \sqrt{R^{2} + \left(
        b + \sqrt{z^{2}+c^{2}} \right)^{2}}},
\end{eqnarray}
with $M_{\rm d} = 10^{11}$~M$_{\odot}$, $b = 6.5$~kpc and $c =
0.26$~kpc. Finally, the bulge is modelled as a Hernquist potential
\begin{eqnarray}
  \label{eq:bulgepot}
  \Phi_{\rm bulge}(r) & = & \frac{G M_{\rm b}} {r+a},
\end{eqnarray}
using $M_{\rm b} = 3.4 \times 10^{10}$~M$_{\odot}$ and $a=0.7$~kpc.
The superposition of these components gives quite a good
representation of the Milky Way. The circular speed at the solar
radius is $\sim 220$~km\,s$^{-1}$.  The major advantage is the
analytical accessibility of all quantities (forces, densities, and so
on).  Hence, this model has been very widely used -- in particular, in
many previous investigations of the Sgr
stream~\citep[e.g.][]{He04a,He04b,Jo05,La05}.

In the second (denoted by DB), we use the Galactic potential suggested
by \citet{De98}. It consists of three disc components, namely the
ISM, the thin and the thick disc, each of the form
\begin{eqnarray}
  \label{eq:dehndisc}
  \rho_{\rm disc}(R,z) & = & \frac{\Sigma_{\rm d}}{2z_{\rm d}} \exp
  \left( - \frac{R_{\rm m}}{R} - \frac{R}{R_{\rm d}} -
      \frac{|z|}{z_{\rm d}} \right).
\end{eqnarray}
With $R_{\rm m}=0$, equation~(\ref{eq:dehndisc}) describes a standard
double exponential disc with scale-length $R_{\rm d}$, scale-height
$z_{\rm d}$ and central surface-density $\Sigma_{\rm d}$.  For the
stellar disks $R_{\rm m}$ is set to zero, while for the ISM disc, we
allow for a central depression by setting $R_{\rm m}=4$~kpc.
Furthermore, the halo and the bulge are represented by two spherical
density distributions of the form
\begin{eqnarray}
  \label{eq:dehnspher}
  \rho_{\rm S}(R,z) & = & \rho_{0} \left( \frac{m}{r_{0}}
  \right)^{-\gamma} \left( 1 + \frac{m}{r_{0}} \right)^{\gamma -
    \beta} \exp \left( - \frac{m^{2}} {r_{\rm t}^{2}} \right),
\end{eqnarray}
where $m^2 = R^2 + z^2 q^{-2}$ and $q$ is the axis ratio in the
density. We choose the parameters according to the best-fit model 4 in
\citet{De98}. This provides a better representation of the Galaxy, but
at somewhat greater computational cost.

\subsection{Results}

Snapshots of the distribution of tidal debris around the North
Galactic Cap for some typical simulations are shown in
Figure~\ref{fig:sims}.  We quickly see that most of the models do not
look at all like the data.  Only haloes close to spherical provide
bifurcated streams.  The simulated streams in moderately and strongly
oblate or prolate haloes do not bifurcate.

To proceed further, we need to develop an objective criterion for
identifying the bifurcation. We use a Marquand-Levenberg routine to
fit a single Gaussian and two Gaussians to the declination
distribution in the young leading and old trailing tidal
debris. Models for which a single Gaussian is everywhere preferred (as
judged by the $\Delta \chi^2$) are unacceptable, as they do not show
two identifiable streams. If two Gaussians are a better fit than a
single, then the ratio $d$ of the distance between the two peaks to
the sum of the dispersions of each peak is computed as a function of
right ascension.  We refer to the mean value of this parameter, taken
over all right ascensions $110^\circ \le \alpha \le 220^\circ$, as the
strength of the bifurcation, $\langle d \rangle$. The onset of the
bifurcation $\alpha_0$ is taken to be the right ascension when $d =
1.5$.  Applying this algorithm to Belokurov et al.'s (2006) dataset,
the bifurcation has strength $\langle d \rangle \approx 1.7$ and
begins at a right ascension $\alpha_0 \approx 190^\circ$.  For a range
of simulations, the same quantities are recorded in
Table~\ref{tab:simres}.  Both Galaxy models contain a flattened disk
and bulge, so the model parameters ($q_\phi$ and $q$) are not a
reliable guide to the overall flattening of the potential. Rather, we
give in Table~\ref{tab:simres} the axis ratio of the equipotentials
$Q_\phi$ at the mean of the pericentric and apocentric distances of
the Sgr's orbit.

There are a number of interesting conclusions from the Table.  First,
very few models actually give bifurcations at all.  Of the 80 models
tested, only 10 give bifurcated streams.  A bifurcation occurs if the
axis ratio of the potential $Q_\phi$ lies in the range $0.92 \lesssim
Q_\phi \lesssim 0.97$. The best overall match to the data is given by
simulations using the Miyamoto-Nagai disk and logarithmic halo with
$q_\phi=1$, together with sets c or d of proper motions. They
reproduce the strength and the location of the bifurcation reasonably
well.  Second, if we use the proper motions measured by HST \citep[set
a, derived by][]{Ib01b}, we do not obtain bifurcated streams, whatever
the halo flattening.  The precession is controlled not just by the
flattening of the potential, but also by the eccentricity of the
orbit, and hence the proper motions.  Sets b and c of proper motions,
with values close to those derived from Schmidt plates \citep{Ir96},
provide much better fits.  Third, \citet{He04b} has claimed that
strongly prolate haloes with $q \approx 1.65$ give the best fit to the
2MASS data on the right ascension, declination, heliocentric distance
and radial velocity of the tidal debris. Such strongly prolate models
do not match the bifurcated stream in the SDSS dataset. In fact, the
claims of prolateness rely heavily on their choice of Galactic
potential (Miyamoto-Nagai disk and logarithmic halo) and are {\it not}
reproduced with the Dehnen \& Binney models. The Miyamoto-Nagai disk
declines like a power-law, rather than an exponential, and so it is
reasonable to interpret the finding of prolateness or stretching of
the halo as compensation for deficiencies in the disk model.
Similarly, although some of the nominally prolate halo models
($q_\phi>1$) in Table~\ref{tab:simres} provide matches to the
bifurcation data, the equipotentials are mildly oblate ($Q_\phi <1$)
at the radii probed by Sgr's orbit.

Nearly spherical models can reproduce the projected density of the Sgr
stream around the Northern Galactic Cap.  However, as is traditional
in this area, our simulations do not fit all the data! For example,
there is a mismatch of $\sim 20^\circ$ in the right ascension of the
beginning of the C stream (c.f., the upper and lower right panels of
Figure~\ref{fig:xz}). Figure~\ref{fig:error} shows the heliocentric
distances and velocities of stream A and B for the simulation of
Figure~\ref{fig:xz}, together with the heliocentric distances derived
by Belokurov et al. (2006) from subgiant branch fitting.  The
heliocentric distances of the simulated streams are within the
observational error bars over the range of right ascensions $\alpha
\gtrsim 190^\circ$, but they are too small for $\alpha \lesssim
190^\circ$. In common with other simulations in oblate haloes
\citep[see e.g.,][]{Jo05}, the radial velocities of 2MASS M giants in
the leading arm are also not matched. Possible causes of these
discrepancies are discussed shortly.

\begin{figure}
\begin{center}
\includegraphics[height=13cm]{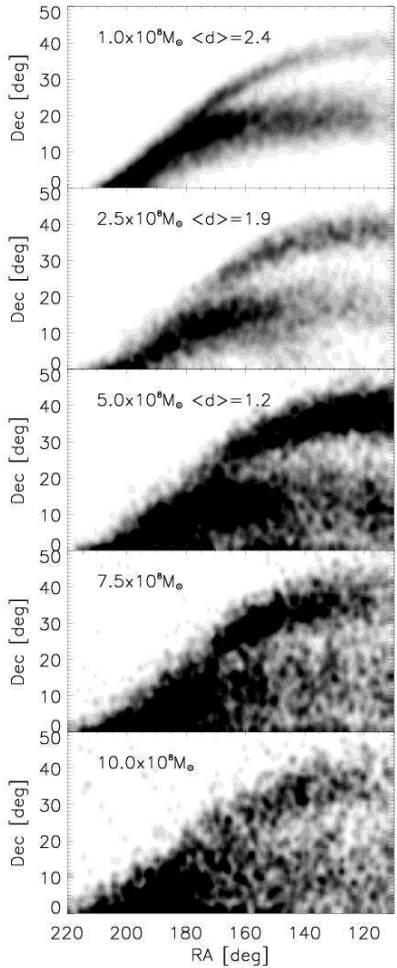}
\caption{\label{fig:masssequence} A sequence of simulations differing
only in the mass of the Sgr dSph at the beginning of our 10 Gyr
simulations.  This is marked in the top left-hand corner and varies
from $2.5 \times 10^8 \Msun$ (top panel) to $10^9 \Msun$ (bottom
panel). If there is a bifurcation, then strength $\langle d \rangle$
is also recorded.  The Galactic model is a Miyamoto-Nagai disk and
logarithmic halo with $q_\phi = 1.05$, while proper motion set d is
used.}
\end{center}
\end{figure}
\begin{table}
\begin{center}
\caption{\label{tab:sagdata} Properties of the Models of the Sgr dSph.}
\begin{tabular}{cccccc}
\hline
Initial Mass & Final Mass & Final Velocity & Final Crossing \\  
(in $10^8\,\Msun$) & (in $10^8\,\Msun$) & Dispersion (in km\,s$^{-1}$) & Time (in Myr)\\ 
\hline
1.  & 0.5 & 19. & 59 \\
2.5 & 1.9 & 30. & 38 \\
5.  & 3.4 & 36. & 46 \\ 
7.5 & 5.7 & 44. & 37 \\
10 &  8.0 & 50. & 32 \\\hline
\end{tabular}
\end{center}
\end{table}

The existence of the bifurcation also constrains the mass of the Sgr.
For example, Figure~\ref{fig:moremass} shows a simulation identical to
that of Figure~\ref{fig:xz}, but with the initial mass of the Sgr
increased to $5 \times 10^8 M_\odot$. The more massive the Sgr, the
greater its internal velocity dispersion. This has two consequences --
the tidal arms are broader and they diffuse away from the Sgr's
orbital path more quickly. The overall effect is that debris is
scattered over a wider range of locations. As the subpanels on the
right hand side show, the bifurcation persists but is less dramatic
than in the data, whilst the A and B streams are no longer as
collimated as in the data. The run of heliocentric distances with
right ascensions, however, is a somewhat better match.
Figure~\ref{fig:masssequence} shows a sequence of simulations with Sgr
masses between $10^8 \msun$ and $10^9 \msun$, while the Galactic
potential is kept fixed as a Miyamoto-Nagai disk and logarithmic halo
with $q_\phi = 1.05$. The bifurcation blurs with increasing
mass. Quantitively, the strength of the bifurcation $\langle d
\rangle$ falls from 2.4 when the Sgr's mass is $10^8 \msun$ to 1.9 at
$2.5 \times 10^8 \msun$ and to 1.2 at $5 \times 10^8 \msun$. Once the
Sgr mass rises much above $5 \times 10^8 \msun$, there is no visible
bifurcation.  High mass models are disfavored until it has been
demonstrated that the tidal streams of stars can remain as highly
collimated as in the data. In Table~\ref{tab:sagdata}, we give the
properties of the Sgr remnant at the end of our simulations.  Of
course, our simulations show both dark matter particles and stars,
whereas the observable today is the luminous matter left in the Sgr
dSph. The mass loss of the Sgr is mainly affected by its orbit and
hence the choice of proper motion together with the choice of Galactic
potential. Since these are uncertain, we also give in
Table~\ref{tab:sagdata} the internal crossing times at the virial
radius and the three dimensional velocity dispersion, which governs
the broadening of the tails.

Our overall picture gives three predictions. First, a second wrap of
branch D may be detectable in the 2MASS data, closer than that already
reported by \citet{Ma03}.  Secondly, the dynamically older B stream
should have a larger velocity dispersion than the younger A stream.
The mean velocities of branches A and B probably differ by $\sim 15$
kms$^{-1}$, though this may be hard to measure as it may be less than
the stream's internal dispersions. Thirdly, if our current ideas on
the star formation history of the Sgr are
correct~\citep[e.g.,][]{Gr00}, then there may be a difference in the
stellar populations of the streams. Stream A may contain evidence for
a younger population that is not present in stream B.

\section{Conclusions}

The Sgr stream, as seen by the Sloan Digital Sky Survey
(Belokurov et al. 2006), is composed of two branches (A and B) at
about the same heliocentric distance visibly diverging at right
ascensions $\alpha \approx 190^\circ$ to give a bifurcation, together
with a third stream (C) aligned with the A branch, but well behind
it. This complex and intricate morphology throws down an enormous
challenge to modellers.  

Here, we have given a physical picture of how this structure may
arise. The bifurcation is caused by a projection of the young leading
(A) and the old trailing (B) tidal arms of the Sgr, while the old
leading arm (C) lies well behind A.  The bifurcation between A and B,
and the positioning of C behind A, can only be reproduced in
simulations of the disruption of the Sgr if the halo is nearly
spherical.  Simulations in either moderately or strongly oblate or
prolate haloes fail these tests by a wide margin. In particular, the
bifurcation only exists if the axis ratio of the potential $Q_\phi$ at
the radii sampled by the Sgr's orbit lies in the range $0.92 \lesssim
Q_\phi \lesssim 0.97$. The physical explanation of this is easy to
provide. The material in the A and B branches is about two revolutions
apart in orbital phase. The angular difference on the sky between the
A and B branches is therefore a direct measure of the precession of
the orbital plane of the Sgr over two revolutions. As the angular
difference is small, so the precession of the orbital plane is small,
and so the potential muss to be close to spherical. If the potential
is moderately prolate or oblate, debris is scattered over a much wider
range of locations.  The path from observation to theoretical
conclusion is surprisingly direct and independent of detailed
modelling.

The bifurcation also provides a strong constraint on the mass of the
Sgr and its debris. If this is much larger than $5 \times 10^8 \msun$,
then the tidal streams are too diffuse to give a clear bifurcation.
This also is easy to understand. As the internal velocity dispersion
increases with progenitor mass, so the streams become broader and
diffuse more quickly in the Galactic potential. The A and B branches
do not then have the highly collimated appearance seen in the data.

Our simulations -- like all other simulations of the disruption of the
Sgr in the literature -- do not agree with all the data. In
particular, the detailed distances to the A, B and C streams given in
Belokurov et al. (2006) are not reproduced over the full range of
right ascension. Although this is a defect, the limitations of the
commonly-used methodology for Sgr disruption simulations also need to
be acknowledged. The underlying assumption is that the Galactic
potential is static and unevolving over up to 10 Gyr timescales. This
is clearly incorrect -- the Milky Way is believed to have accreted $30
\%$ of its mass over the last 5 Gyr~\citep[see
e.g.,][]{Bo02,Ne06}. Simulations typically show that the last major
merger take place at about a redshift $z=1$, roughly 8 Gyrs
ago~\citep{NFW95}. Although most of the mass will have accreted in the
outer parts of the Galaxy, the Sgr's orbit extends out to $\sim 60$
kpc and will surely been affected by this rearrangement. The formation
of the Galactic bar has been dated to between 5 and 8 Gyr
ago~\citep[see e.g.,][]{Se99a,Se99b}, and so bar-driven evolution of
the inner Galaxy will also have caused substantial changes.  The
effects of time evolution are of much greater importance for the SDSS
dataset than for the 2MASS dataset, which is restricted to dynamically
younger material.

The strength of the argument presented in this paper is that it relies
on the gross morphological features of the Sgr stream. To reproduce
the detailed positions and velocities of stars in the A, B and C
branches may well require a clearer understand of Galactic
evolution. However, the existence of a bifurcation in nearly spherical
potentials is a robust result. To challenge the main conclusion of
this paper requires the devising of an alternative explanation of the
existence of two streams that are closely matched in distance over a
$\sim 50^\circ$ arc.  In this respect, our argument compares
favourably with other methods of determination of halo shape using
methods such as the flaring of the neutral gas
layer~\citep[e.g.][]{Ol00} or the stellar kinematics of halo
stars~\citep[e.g.,][]{Ma91}. These are afflicted by systematic
uncertainties regarding the contribution of the cosmic ray pressure
or the orientation of the stellar velocity ellipsoid, for example. In
contrast, the bifurcation in the Sgr stream is a clean, simple and
direct test.

\acknowledgments
\noindent
MF, VB, MIW and DZ thank the Particle Physics and Astronomy Research
Council of the United Kingdom for financial support, and Walter Dehnen
for providing software. We thank Eric Bell, Heidi-Jo Newberg, Connie
Rockosi, Brain Yanny and Kathryn Johnston (the referee) for
constructive comments on the draft version. Funding for the SDSS and
SDSS-II has been provided by the Alfred P.  Sloan Foundation, the
Participating Institutions, the National Science Foundation, the
U.S. Department of Energy, the National Aeronautics and Space
Administration, the Japanese Monbukagakusho, the Max Planck Society,
and the Higher Education Funding Council for England. The SDSS Web
Site is http://www.sdss.org/.
                                 
The SDSS is managed by the Astrophysical Research Consortium for the
Participating Institutions. The Participating Institutions are the
American Museum of Natural History, Astrophysical Institute Potsdam,
University of Basel, Cambridge University, Case Western Reserve
University, University of Chicago, Drexel University, Fermilab, the
Institute for Advanced Study, the Japan Participation Group, Johns
Hopkins University, the Joint Institute for Nuclear Astrophysics, the
Kavli Institute for Particle Astrophysics and Cosmology, the Korean
Scientist Group, the Chinese Academy of Sciences (LAMOST), Los Alamos
National Laboratory, the Max-Planck-Institute for Astronomy (MPIA), the
Max-Planck-Institute for Astrophysics (MPA), New Mexico State
University, Ohio State University, University of Pittsburgh,
University of Portsmouth, Princeton University, the United States
Naval Observatory, and the University of Washington.

\end{document}